# AI-Generated 3D Environments as Speculative Mediators in More-Than-Human Design: An Exploratory Study

Aung Pyae*

International School of Engineering, Faculty of Engineering, Chulalongkorn University, Bangkok, Thailand, aung.p@chula.ac.th

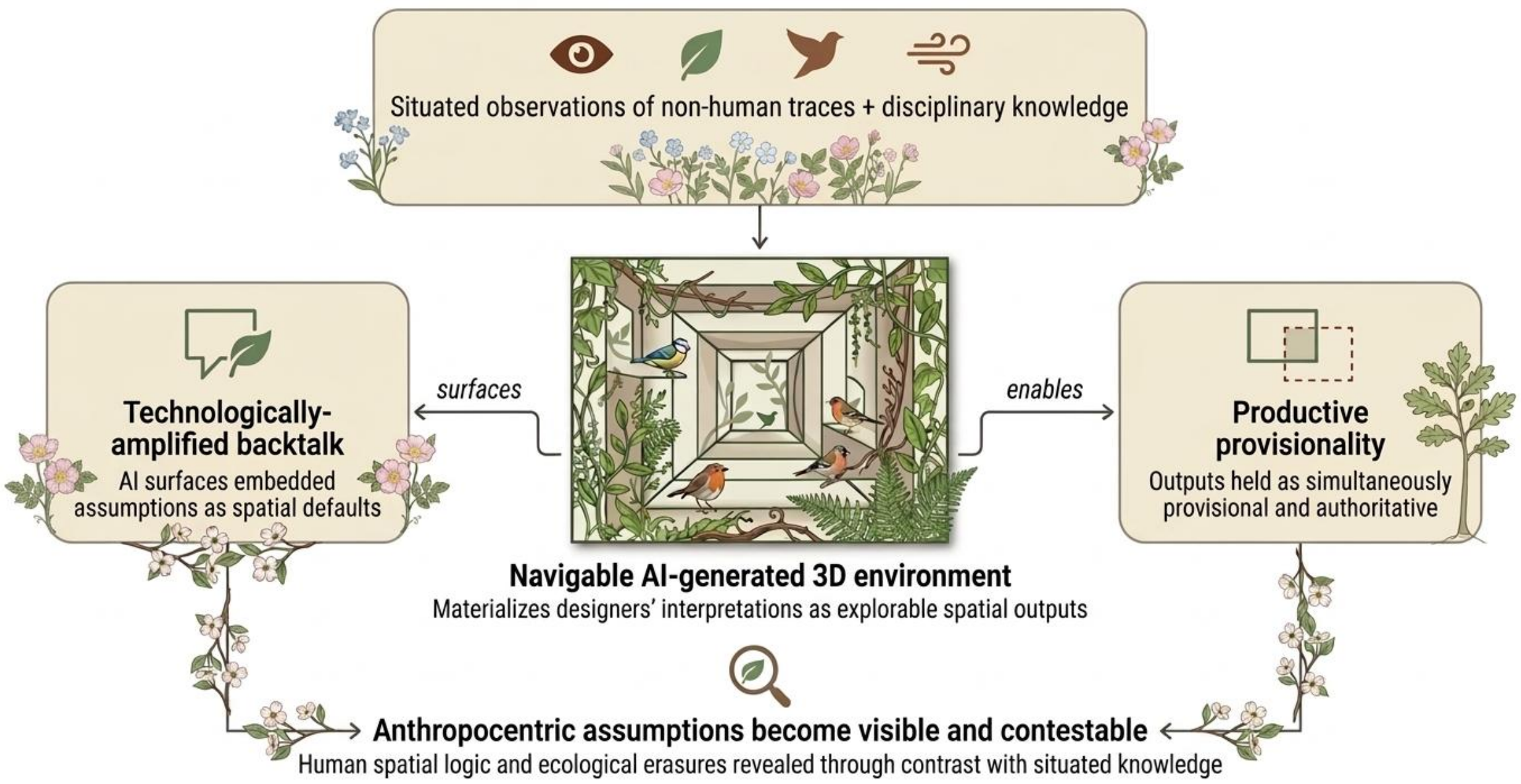

Figure 1: Navigable AI-generated environments surfacing anthropocentric assumptions in more-than-human design

More-than-human design challenges anthropocentric assumptions by foregrounding non-human entities as stakeholders, yet designers face an epistemic boundary: they cannot directly access non-human experience. We present an exploratory study examining how generative AI—specifically a text-to-3D world generation platform producing navigable environments—may function as a speculative mediator in more-than-human design. Through a qualitative study with five participants from engineering and sustainability backgrounds engaging with AI-generated worlds derived from non-human traces, we investigate how instant exploration—navigating generated environments within seconds—shapes reflection, iteration, and provisional treatment of outputs. Our findings suggest that navigating AI-generated environments supports reflection-in-action distinct from evaluating static representations, while designers' epistemic stances oscillate between treating outputs as generative provocations and as authoritative representations. We propose

* Place the footnote text for the author (if applicable) here.

technologically-amplified backtalk and productive provisionality as preliminary lenses for understanding how navigable AI-generated 3D environments can surface anthropocentric assumptions in more-than-human design.

## 1 INTRODUCTION

Human–Computer Interaction (HCI) has increasingly questioned whether its foundational human-centeredness is adequate for complex contemporary challenges [Giaccardi 2025]. More-than-human perspectives reposition non-human entities—animals, plants, ecosystems, materials, and technologies—as legitimate stakeholders in design [Coulton 2019] [Forlano 2017], drawing on posthumanist philosophy [Braidotti 2013], new materialism [Bennett 2010], and actor-network theory [Latour 2005] to emphasize that humans and non-humans are entangled in webs of interdependence [Haraway 2016]. In parallel, generative AI systems capable of producing three-dimensional environments from natural language raise new questions for design practice, particularly as research suggests that AI assistance shapes creative processes in complex ways [Doshi 2024].

More-than-human design faces a fundamental epistemic challenge: designers cannot directly access non-human experience [Fuchsberger 2023][Nicenboim 2023]. Engagement with non-human concerns is mediated through indirect traces—scientific data, behavioral observations, and interpretive frameworks [Nicenboim 2023]. Speculative design provides resources for engaging beyond direct experience [Dunne 2013] [Coulton 2017], and recent work has explored AI-supported more-than-human design strategies [Nicenboim 2024], yet the role of AI-generated navigable 3D environments in this context remains underexamined. World generation platforms—systems translating textual descriptions into explorable three-dimensional spaces—offer a distinctive capability: the rapid transformation of conceptual descriptions into environments that can be spatially navigated rather than merely viewed.

This paper presents an exploratory study examining how generative AI—specifically a text-to-3D world generation platform—may function as a speculative mediator in more-than-human design: not as a proxy for non-human entities, but as a tool translating designers' interpretations of non-human concerns into navigable speculative environments. Building on Schön's notion of design [Schön 1983] as a "reflective conversation with the materials of a design situation," we investigate how instant exploration—near-immediate generation and navigation of 3D environments—shapes reflective engagement with non-human concerns. In this study, we ask:

- RQ1 How do designers translate indirect traces of non-human concerns into AI prompts, and what interpretive transformations occur?
- RQ2 How does navigating AI-generated 3D environments shape reflection and iterative refinement?
- RQ3 When and why are AI-generated worlds treated as provisional provocations versus authoritative representations?
- RQ4 How does engagement with AI-generated speculative environments surface or challenge anthropocentric assumptions?

We contribute: (1) two emergent concepts—technologically-amplified backtalk and productive provisionality—as provisional lenses for understanding how navigable AI-generated environments shape reflective engagement with non-human concerns; (2) preliminary empirical insight into how instant spatial exploration creates a distinctive temporality for reflection-in-action in AI-generated 3D environments; and (3) initial design implications for scaffolding epistemic uncertainty in AI-mediated more-than-human speculation, including strategies for preserving productive tension during observation-to-prompt translation.

## 2 RELATED WORK

### 2.1 More-Than-Human Design and Speculative Practice

The more-than-human turn in HCI departs from human-centered design, arguing that anthropocentric assumptions are inadequate for addressing entanglements of humans, non-human organisms, ecosystems, and technologies [Giaccardi 2025]. [Nicenboim 2023] articulate "decentering" as a core practice—deliberate shifts of attention toward non-human needs. Animal-computer interaction [Mancini 2011] [Westerlaken 2016], urban ecology research [Clarke 2018], and "thing ethnography" [Giaccardi 2016] have developed varied approaches, yet all face the epistemic boundary motivating our study. Speculative design uses designed artifacts to explore alternative possibilities [Dunne 2013], with design fiction emphasizing world-building as a mode of speculation [Coulton 2017]. [Lindley 2023] propose "productive oscillation"—deliberately moving between perspectives to avoid claiming a single authoritative viewpoint. We examine where such oscillation occurs in AI-mediated practice, and how it manifests specifically in designers' epistemic stances toward AI-generated outputs.

### 2.2 Reflective Practice and Design Cognition

Schön's reflective practice characterizes design [Schön 1983] as a "reflective conversation with the materials of a design situation," with "reflection-in-action" occurring within the flow of activity. [Goldschmidt 2003] elaborates how external representations "talk back" to designers, revealing unintended patterns—foundational to our exploration of how AI-generated navigable environments may amplify such backtalk through spatial materialization. [Dorst 2001] describe co-evolution of problem and solution, a dynamic potentially intensified when AI enables rapid environmental generation. Design problems have been characterized as "wicked" [Buchanan 1992]—requiring sophisticated judgment particularly crucial when engaging with non-human perspectives.

### 2.3 Generative AI in Design

AI assistance enhances individual creative performance while reducing collective output diversity [Doshi 2024]. [Liu 2023] demonstrate that text-to-image AI can support ideation while anchoring designers to particular representations. [Nicenboim 2024] developed AI tools—"Oblique" and "MoTH"—using large language models to create design strategies from more-than-human texts. Our study extends this line of inquiry from text-based strategies to navigable 3D world generation. While prior work has established that fictional and speculative resources can support perspective-taking in design [Blythe 2006], AI-generated environments introduce a distinctive quality: designers encounter outputs shaped by the model's training defaults and biases rather than by their own deliberate craft, creating conditions where assumptions may be surfaced through contrast. We examine how spatial, walkable environments shape reflective engagement differently from static or text-based outputs. Questions of trust calibration are critical: [Parasuraman 1997] distinguish calibrated from miscalibrated trust in automation, and users may attribute greater authority to spatially immersive AI outputs. Together, these perspectives frame our investigation of how navigable AI-generated 3D environments may shape reflective engagement with non-human concerns, which we examine through an exploratory qualitative study.

## 3 METHOD

We conducted an exploratory qualitative study [Stebbins 2001] using a 90-minute workshop combining situated observation, design thinking, AI-mediated externalization, and reflective interviews. Five participants (aged 24–32) were recruited through purposive sampling from Chulalongkorn University in Thailand: three civil engineering graduate students, one geologist, and one sustainability researcher. We selected participants whose training foregrounds material and environmental systems, providing relevant domain knowledge while remaining unencumbered by speculative design conventions. All had basic AI familiarity (e.g., ChatGPT, Claude, Gemini) but no prior 3D world generation experience. The study followed institutional ethical guidelines with informed consent (P1–P5). Participants used Marble by World Labs, a text-to-3D platform that generates navigable environments from natural language prompts. Unlike text-to-image tools generating static representations, Marble produces explorable three-dimensional spaces that users can walk through and view from multiple perspectives in real time, underpinning our notion of instant exploration.

The workshop comprised five stages: (1) introductory briefing on more-than-human design; (2) a site walk on an elevated campus bridge (see Figure 2), where participants collected traces of human and non-human presence—including bird and insect feces on railings, encroaching vegetation, structural gaps in flooring, and intense heat from an uninsulated roof; (3) individual reflective notes, inspired by Tsing's arts of noticing [Tsing 2015], foregrounding uncertainty and attention to overlooked non-human presences; (4) speculative renovation ideas treating humans, animals, plants, wind, and light as co-present, avoiding optimization language to foreground the wicked nature of multispecies cohabitation [Buchanan 1992]; and (5) AI externalization, where participants translated their design ideas into natural language prompts and generated navigable 3D environments through Marble (see Figure 3 & 4). Participants iteratively refined prompts over one to four generations, responding to mismatches between intentions and AI outputs. The facilitator framed AI as a speculative mirror rather than a designer or validator, and the process concluded with semi-structured interviews. Nine strategically developed and piloted questions scaffolded the workshop, guiding participants from observation through prompt construction to critical reflection. We acknowledge that this structured framing likely influenced participants' strategies; therefore, findings are treated as situated rather than representative.

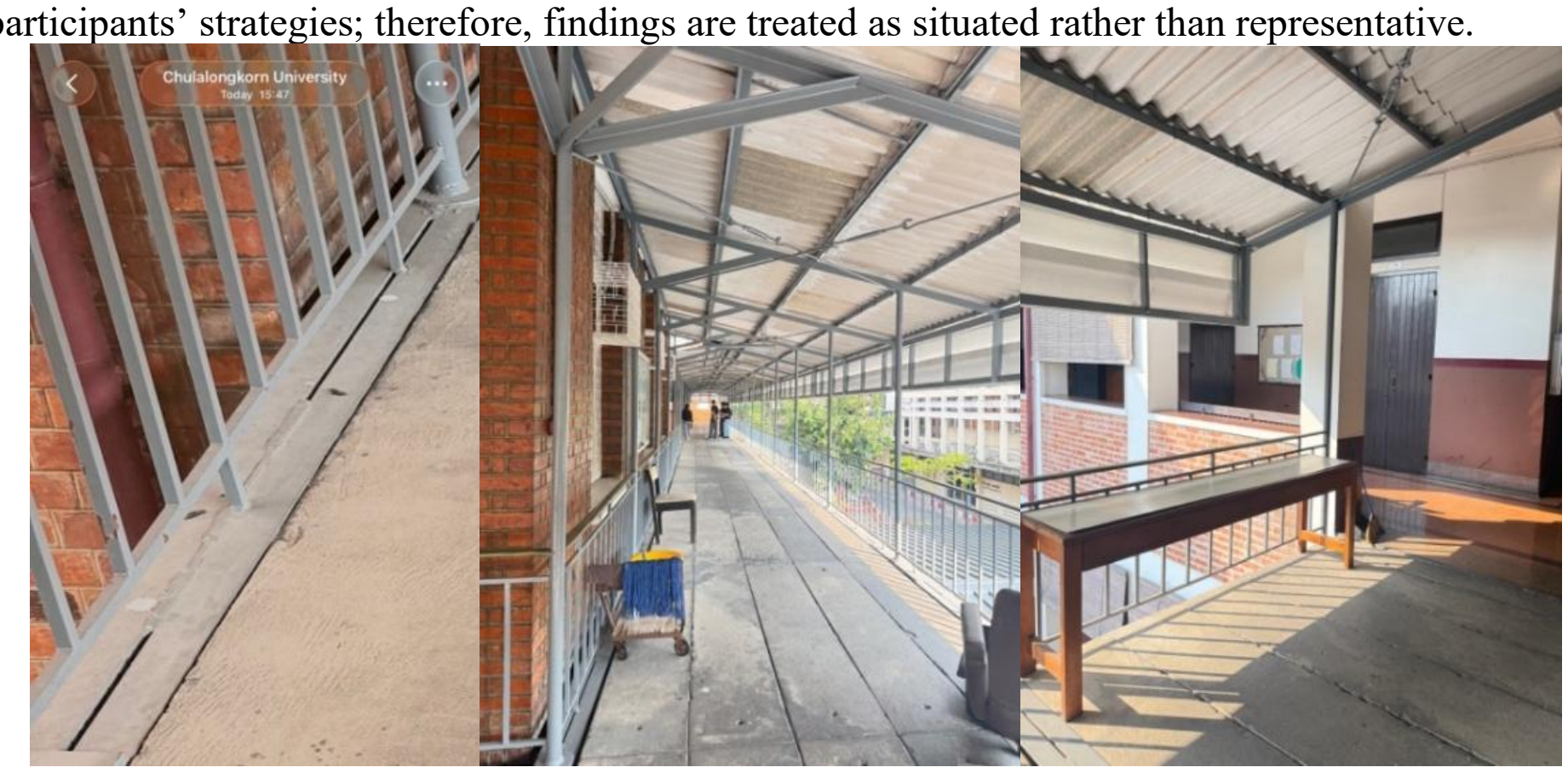

Figure 2: Original bridge photos

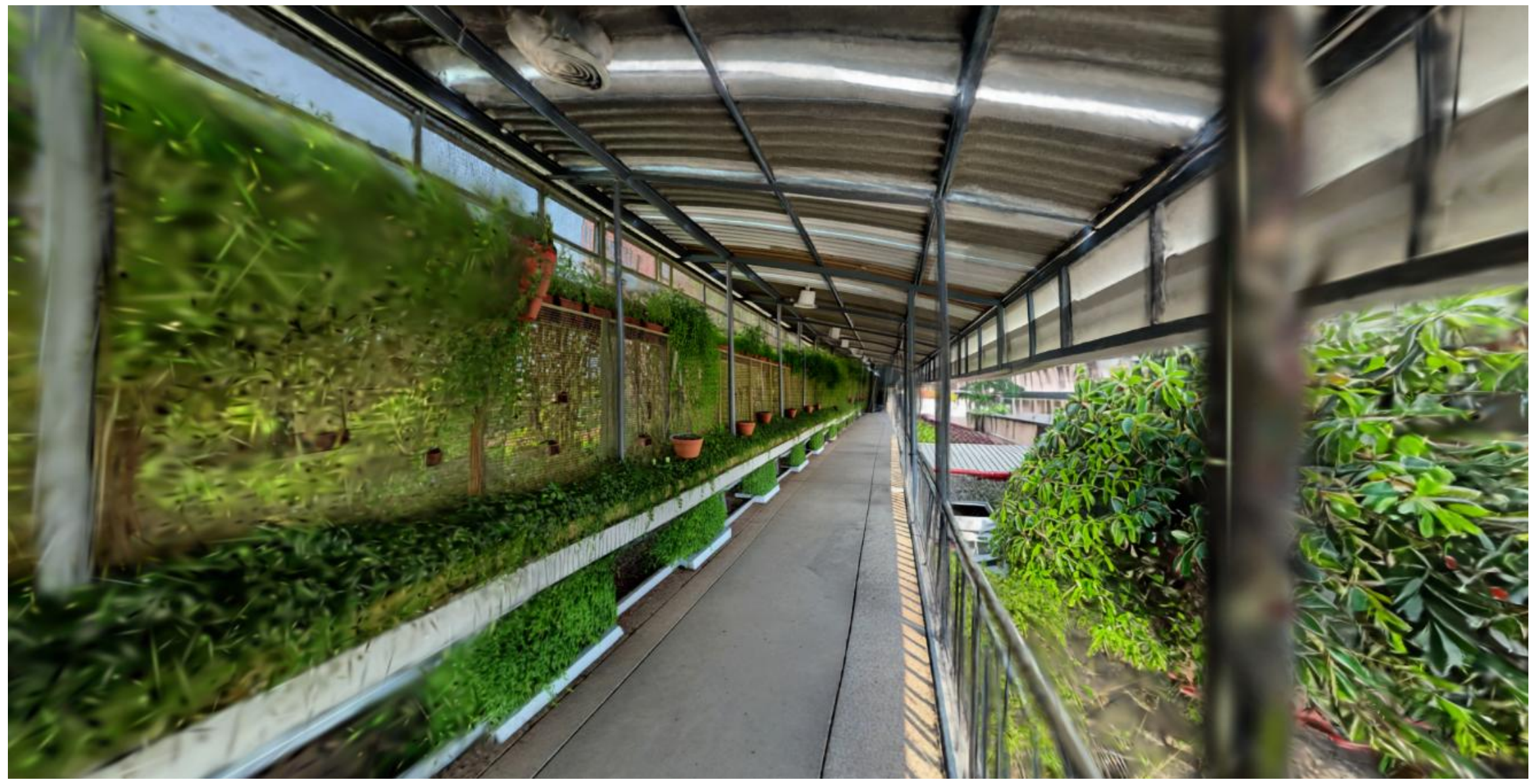

Figure 3: AI-Generated speculative 3D world and design - 1

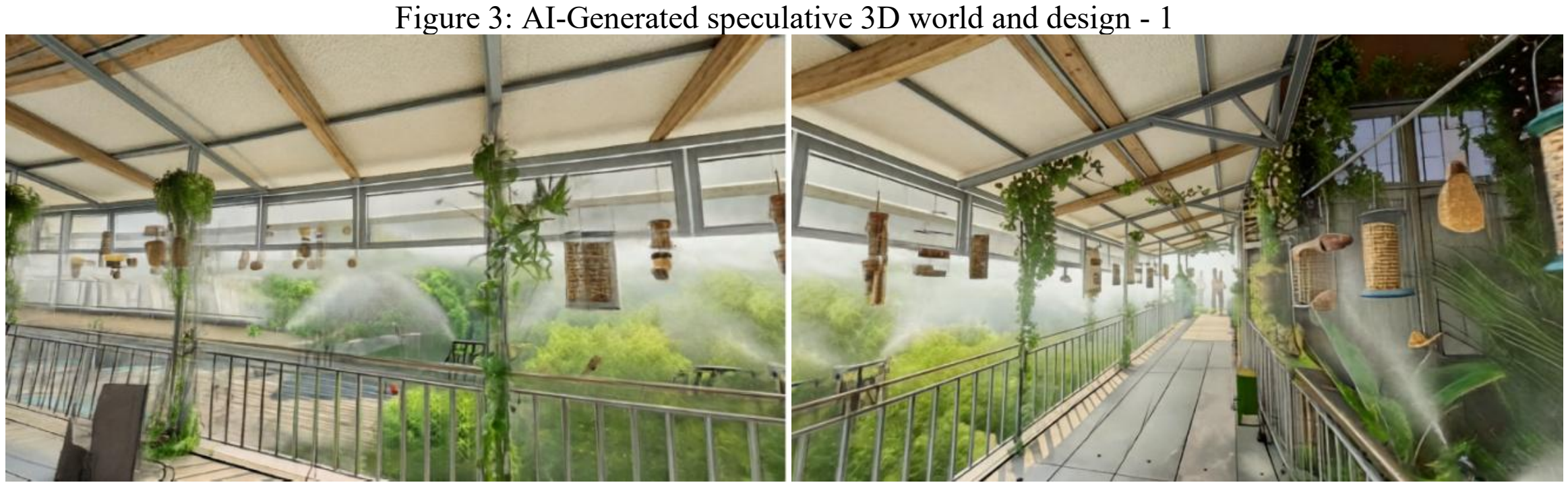

Figure 4: AI-Generated speculative 3D world and design - 2

Data included audio-recorded interviews, written notes, photographic traces, screenshots with prompt histories, and field notes. Analysis used reflexive thematic analysis [Braun 2022], focusing on interpretive transformations during prompt construction, reflective responses during navigation, and epistemic stances toward AI-generated outputs. Prompt histories were triangulated against interview accounts. As an exploratory study, we do not claim saturation or transferability but aim to surface provisional patterns warranting further investigation [Stebbins 2001].

## 4 RESULTS AND FINDINGS

### 4.1 From Observation to Prompt: Interpretive Transformations

During the site walk, all five participants identified concrete non-human traces: bird droppings on railings, encroaching tree branches, insect activity near lighting, and wind patterns. When translating observations into prompts, participants consistently shifted from describing what they observed to prescribing what they envisioned. P1's prompts evolved across four iterations from situated description toward solution-oriented language specifying a green ecosystem focused on animals. P2 introduced precise engineering details absent from her observations—such as timber–concrete composite decking and bird nesting cavities—drawing on structural engineering expertise. P3 iterated across three prompts but noted that elaboration introduced assumptions; despite emphasizing lightweight construction, the AI consistently produced lush greenery, revealing how visual sustainability defaults to abundance rather than efficiency. Across participants, prompt construction activated disciplinary knowledge as an interpretive filter, transforming partial and uncertain traces into confident specifications. This observation-to-prompt translation emerged as a critical site where productive oscillation [Lindley 2023] occurs—specifically between situated observation and disciplinary interpretation.

### 4.2 Navigating Generated Worlds: Immediacy, Mismatch, and Reflective Iteration

Instant exploration—generating a navigable world and immediately walking through it—allowed participants to identify mismatches between intention and AI materialization within the flow of activity. While navigating his first environment, P1 observed that non-humans appeared passive and decorative; encountering this hierarchy through movement prompted subsequent refinements toward explicit multispecies language. P2 treated mismatch as a reflective resource, attributing gaps to her own communicative framing rather than AI limitations. Two reflective strategies emerged: participants generating multiple worlds (P1: 4; P3: 3; P2: 3) engaged in iterative refinement, while P4 and P5 produced a single world relying on front-loaded specificity—P4 through material research and P5 through detailed spatial framing. Both strategies supported reflection through different temporal rhythms: distributed across generate–navigate–revise cycles or concentrated during prompt composition. Navigability was particularly consequential in the iterative mode, enabling participants to discover spatial relationships—such as the positioning of human walkways relative to animal spaces—only through movement.

### 4.3 Oscillation Between Provisionality and Authority

Participants' epistemic stances oscillated between treating AI-generated environments as speculative provocations and as authoritative design proposals. All five recognized the speculative nature: P1 questioned whether designs meaningfully supported non-humans or merely represented them symbolically; P3 noted the AI produced harmonious environments lacking ecological tension; P5 observed that nature appeared curated rather than wild. Simultaneously, participants used generated environments as reasoning anchors—P2 reported relying on the navigable bridge to make decisions, while P3 noted that although outputs diverged from observation, they aligned with the underlying idea, revealing a dual evaluation frame grounded in both empirical observation and conceptual aspiration. Spatial navigation intensified this oscillation: walking through generated environments increased their perceived authority, while movement simultaneously exposed discrepancies with situated knowledge. We tentatively characterize this as productive provisionality: an epistemic stance in which unresolved tension between provisional and authoritative readings supports critical reflection. While related to Lindley et al.'s productive oscillation [Lindley 2023]—which describes deliberate movement between perspectives as a design strategy—productive provisionality refers specifically to the epistemic relationship designers adopt toward AI-generated outputs, holding them simultaneously as provisional provocations and as reasoning anchors.

### 4.4 Surfacing Assumptions Through Systematic Erasure

AI-generated environments surfaced anthropocentric assumptions through both what they depicted and what they systematically erased. Despite explicit prompts for multispecies equity, outputs consistently centered human spatial logic: P1 observed non-humans rendered passive and decorative; P5 noted environments maintained a central human walkway with non-humans spatially contained; P3 observed that spatial organization followed human logic—ordered vertical layers, controlled greenery, and clean pathways. Navigating these environments made such hierarchies particularly salient. A consistent pattern was the erasure of labor, maintenance, ecological consequence, and structural uncertainty. P3 noted that AI-generated environments made structural uncertainty invisible; P1 identified tension between visual harmony and ecological complexity; P5 pointed to the absence of unpredictability. This pattern of systematic erasure resonates with Doshi and Hauser's finding that generative AI reduces diversity in collective outputs [Doshi 2024], suggesting a broader tendency toward convergence on resolved, harmonious representations. In the context of more-than-human design, such resolution removes precisely the tensions that decentering practices seek to foreground [Nicenboim 2023]. Through contrast with situated knowledge, these omissions rendered visible the messiness and ongoing labor required for genuine multispecies cohabitation.

## 5 DISCUSSION

This exploratory study examined how generative AI—specifically a text-to-3D world generation platform—may function as a speculative mediator in more-than-human design by materializing designers' interpretations of non-human concerns as navigable 3D environments. Regarding RQ1, the translation from observation to prompt revealed disciplinary knowledge as an interpretive filter, transforming partial traces into confident specifications. This finding extends Lindley et al.'s notion of productive oscillation [Lindley 2023] by suggesting it occurs specifically between situated observation and disciplinary interpretation during prompt construction.

Regarding RQ2, navigating AI-generated environments supported reflection-in-action [Schön 1983] through distinct temporal rhythms. Participants encountered mismatches experientially while moving through generated worlds, creating conditions for what we term technologically-amplified backtalk—building on Goldschmidt's concept of design backtalk [Goldschmidt 2003], where assumptions embedded in prompts were materialized as spatial hierarchies and defaults discoverable through movement. Where Goldschmidt's original formulation describes how sketches reveal unintended patterns to their makers through visual properties, technologically-amplified backtalk operates through a different mechanism: the AI system's training defaults and resolution tendencies produce spatial materializations that talk back not only through visual form but through navigable spatial relationships—walkways, boundaries, hierarchies—that embody assumptions discoverable only through embodied movement.

The amplification here is specific: the speed of generation, the navigability of outputs, and the AI's tendency to resolve ambiguity into coherent environments together produce backtalk that is both more immediate and more spatially legible than that of static representations. Two reflective strategies emerged — iterative refinement through generate–navigate–revise cycles and front-loaded specificity during prompt composition — suggesting that technologically-amplified backtalk operates through different temporal structures depending on designers' engagement strategies.

For RQ3, participants adopted dual evaluation frames, simultaneously treating outputs as provocations and reasoning anchors. This oscillation is more nuanced than simple over-trust or under-trust [Parasuraman 1997], as designers assessed outputs against both empirical observation and conceptual aspiration. We tentatively characterize this stance as productive provisionality, in which unresolved tension between provisional and authoritative readings supports critical reflection. This stance also extends Parasuraman and Riley's [Parasuraman 1997] framework beyond calibrated trust toward a productive use of miscalibration itself: rather than seeking appropriate trust levels, productive provisionality leverages the unresolved tension between over-trust and under-trust as a resource for surfacing assumptions that might otherwise remain invisible.

Addressing RQ4, AI-generated environments surfaced anthropocentric assumptions through systematic erasure—centering human spatial logic and erasing labor, maintenance, and uncertainty despite explicit prompts for multispecies equity. This pattern resonates with Doshi and Hauser's finding that generative AI reduces diversity in collective outputs [Doshi 2024], suggesting a broader tendency toward convergence on resolved representations that removes precisely the tensions decentering practices seek to foreground [Nicenboim 2023]. These erasures, through contrast with situated knowledge, rendered visible the messiness genuine multispecies cohabitation entails.

These preliminary findings suggest that navigable AI-generated 3D environments can surface anthropocentric assumptions not by bridging the epistemic gap with non-human experience, but by making that gap visible through systematic erasure and spatial materialization. These preliminary findings carry implications for AI-supported speculative tools in more-than-human design. First, regarding workshop design: the observation-to-prompt translation emerges as a critical moment where disciplinary knowledge filters non-human traces. Practitioners could scaffold this by requiring designers to articulate what their prompts exclude alongside what they specify—for example, by maintaining an "erasure log" documenting non-human elements observed on-site but absent from prompts. Second, regarding platform design: AI systems could preserve productive uncertainty by generating deliberately incomplete environments—visualizing confidence levels, flagging systematically erased elements, or producing multiple divergent outputs from a single prompt to resist convergence toward resolved harmony. Third, regarding evaluation: the provisionality–authority oscillation suggests value in structured side-by-side comparison between AI-generated environments and situated observations, enabling designers to identify where AI defaults override situated knowledge. Technologically-amplified backtalk and productive provisionality offer preliminary lenses for understanding how designers navigate epistemic uncertainty in AI-mediated more-than-human speculation.

This study has clear limitations. With five participants from a single university, one site, session, and platform, findings are preliminary and not generalizable. Workshop framing likely shaped strategies in ways we cannot fully disentangle. Crucially, our study lacks a comparison between navigable environments and static representations; claims about navigation's distinctive reflective quality rest on participant self-report within a single modality. Additionally, because all participants lacked speculative design experience, we cannot fully distinguish whether observed patterns reflect the AI tool's mediating role or the novelty of speculative practice itself. Findings are also bound to Marble by World Labs' specific affordances and training defaults—observed patterns such as lush greenery defaults and systematic erasure of messiness may reflect this platform's training data rather than text-to-3D generation broadly. Future work should compare navigable 3D generation with other AI modalities, include participants experienced in speculative design, replicate across platforms with different aesthetic defaults, and examine whether productive tensions stabilize or resolve over extended engagements.

In conclusion, navigable AI-generated 3D environments function not as representations of non-human perspectives, but as spatial materializations of designers' own assumptions. Through navigation, these assumptions become visible, contestable, and available for reflection. Technologically-amplified backtalk and productive provisionality offer preliminary lenses for understanding how generative AI supports the epistemic work of more-than-human design—not by overcoming the epistemic boundary between human designers and non-human experience, but by making that boundary productively visible. Generative AI does not help designers access non-human perspectives; it helps them confront the limits of doing so by externalizing their assumptions as navigable, contestable spatial forms. Rather than bringing designers closer to non-human experience, these environments make the impossibility of such closeness productively visible — the systematic erasures revealed by navigation foreground the epistemic boundary itself as a resource for critical reflection.